\documentclass[doublecol]{epl2} 
\usepackage[applemac]{inputenc}
\usepackage[T1]{fontenc} 
 \usepackage{amsmath} 
\usepackage{subfigure}
\usepackage{lipsum}
\usepackage{amsfonts} 
\usepackage{amssymb, mathrsfs}
\usepackage{braket}
\usepackage{graphicx} 

\usepackage{bbm}

\def\beq{\begin{equation}}
\def\eeq{\end{equation}}
\def\bsp{\begin{split}}
\def\esp{\end{split}}
\def\bea{\begin{eqnarray}}
\def\eea{\end{eqnarray}}
\def\ba{\begin{array}}
\def\ea{\end{array}}

\def\lb{\left(}
\def\rb{\right)}

\def\l.{\left.}
\def\r.{\right.}

\title{Antiferromagnetic molecular nanomagnets with odd-numbered coupled spins}
\shorttitle{Antiferromagnetic molecular nanomagnets with odd-numbered coupled spins} 

\author{S. A. OWERRE\inst{1,2\footnote{solomon.akaraka.owerre@umontreal.ca} } \and J. NSOFINI\inst{3,4\footnote{jnsofini@uwaterloo.ca}} }

\shortauthor{S. A. OWERRE and J. NSOFINI}

\institute{                    
  \inst{1} D\'epartement de 
Physique,
Universit\'e de Montr\'eal -
Montr\'eal,  
Qu\'ebec H3C 3J7, Canada\\
  \inst{2} Perimeter Institute for Theoretical Physics - 31 Caroline St. N., Waterloo, Ontario N2L 2Y5, Canada\\
  \inst{3} Institute for Quantum Computing, University of Waterloo - Waterloo, Ontario N2L 3G1, Canada\\
  \inst{4} Department of Physics and Astronomy, University of Waterloo - Waterloo, Ontario N2L 3G1, Canada
}

\abstract{
In recent years, studies on cyclic molecular nanomagnets have captivated the attention of researchers.  These magnets are finite in size and contain very large spins. They are interesting because they possess macroscopic quantum tunneling of N\'eel vectors. For antiferromagnetic molecular nanomagnets  with finite number of even-numbered coupled spins,  tunneling  involves two classical localized N\'eel ground states separated by a magnetic energy barrier. The question is: can such phenomena be observed in nanomagnets with odd number of magnetic ions? The answer is not directly obvious because cyclic chains with odd-numbered coupled spins are frustrated as one cannot obtain a perfect N\'eel order. These frustrated spins can indeed be observed experimentally, so they are of interest.  In this Letter, we theoretically investigate macroscopic quantum tunneling in these odd spin systems with arbitrary spins $s$, in the presence of a  magnetic field applied along the plane of the magnet. In contrast to systems with an even-numbered coupled spins, the ground state of the cyclic odd-spin system contains a  topological soliton due to spin frustration. Thus the classical ground state is $2N$-fold degenerate as the soliton can be placed anywhere along the ring with total $S_z=\pm s$. Small quantum fluctuations delocalize the soliton with  a formation of an energy band. We obtain this energy band using degenerate perturbation theory at order $2s$. We show that the soliton ground state is chiral for half-odd integer spins and non-chiral for integer spins.  From the structure of the energy band we infer that as the value of the spin increases the inelastic polarized neutron-scattering intensity may increase or decrease depending on the strengths of the parameters of the Hamiltonian.}

\begin{document}

\maketitle

\textbf{Introduction} --. 
Quantum antiferromagnetic spin chain is one of the subjects in quantum magnetism endowed with rich and  interesting physics \cite{ dane, bethe1, bin, kit}.   In recent years, a new class of quantum antiferromagnetic spin chain has emerged in molecular wheels, which comprises molecules with an even number of large magnetic spins arranged periodically on a ring \cite{og, ml2001}. These molecular wheels have finite length a few  magnetic ions (sites) coupled antiferromagnetically. This makes their behaviour different from the conventional quantum spin chain \cite{bethe1, dane}, which usually contains small magnetic spins and large number of spin sites (thermodynamic limit). The two remarkable differences between molecular wheels and conventional quantum spin chain are as follows.   $(1).$ Molecular wheels possess macroscopic quantum tunneling of the N\'eel vectors  separated by a magnetic energy barrier. $(2).$ Molecular wheels  possess neither magnetic phase transition nor antiferromagnetic long-range order due to the shortness of the ring (chain) and the presence of very large spin length. However, as the length of the spin decreases and the system goes to the thermodynamic limit, these effects might be observed. Thus, these molecular wheels provide adequate knowledge of quantum and classical regimes.

In the last few years several molecular nanomagnets have been synthesized and studied experimentally by inelastic neutron scattering method \cite{og, og1}.  There have been numerous applications of molecular magnets, however, molecular nanomagnets with odd-numbered coupled spins  pose serious challenges and spin frustration effects might be observed \cite{gatt}. 
There are few odd rings; recently some  odd-numbered magnetic rings have been synthesized \cite{cad} and studied by inelastic neutron scattering and exact numerical diagonalization \cite{cr}. This includes Cr$_7$  with $s = 3/2$ and $N=7$, which has open boundary condition and thus there is no spin frustration.  Numerous studies on molecular nanomagnets have focused on the pure isotropic Heisenberg model \cite{mann, enz,enz1}. This is usually a first approximation because most magnets have dominant exchange interaction. The inclusion of anisotropy and magnetic fields  can  induce exotic phenomena.

\begin{figure}[htd]
\centering
\includegraphics[scale=0.3]{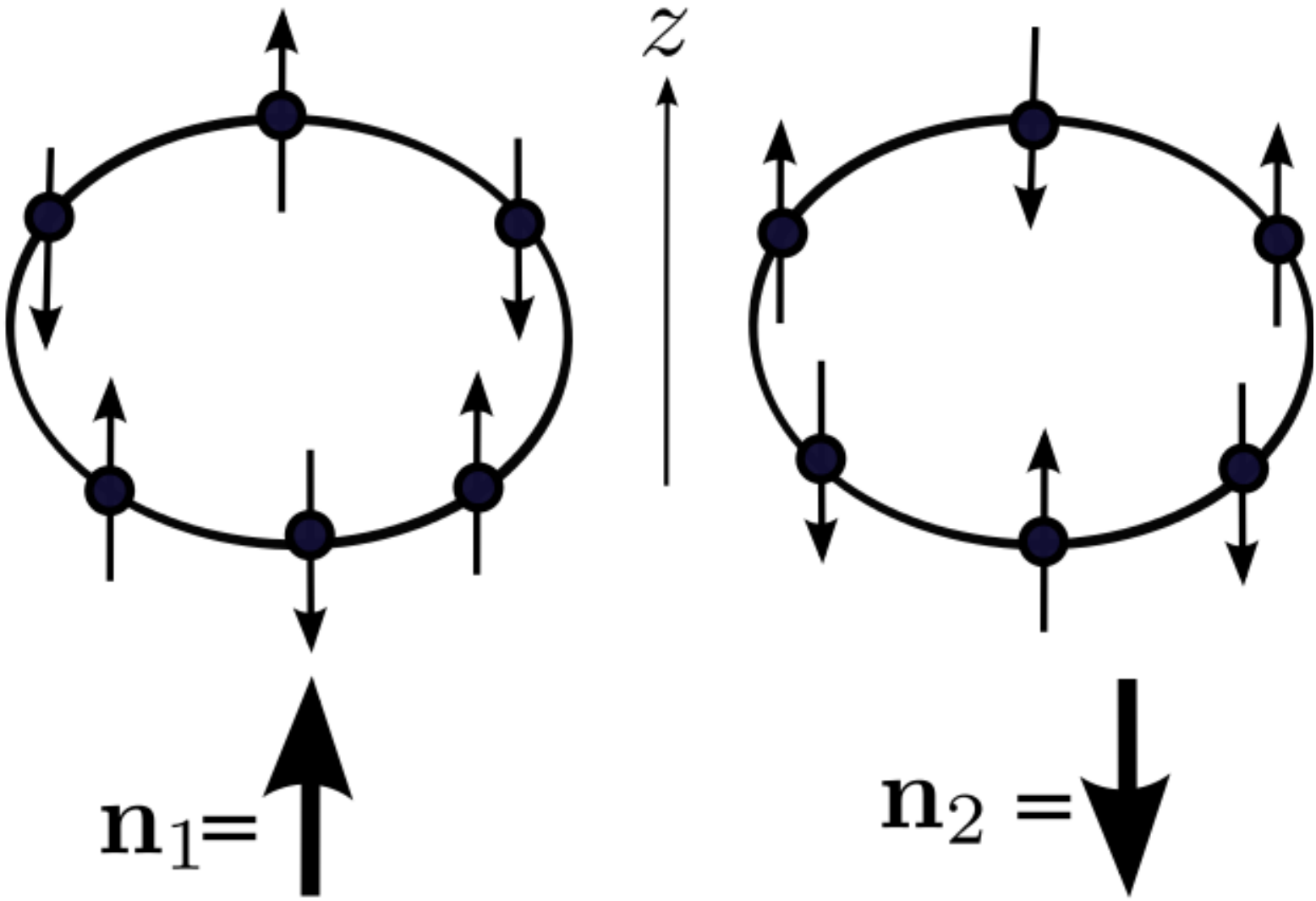}
\caption{Schematic sketch of the two classical degenerate N\'eel ground states for antiferromagnetic molecular nanomagnets with even-numbered coupled spins in zero magnetic field, separated by a magnetic energy barrier of height $2DNs^2$.}
\label{ring}
\end{figure}

\begin{figure}[htd]
\centering
\includegraphics[scale=0.3]{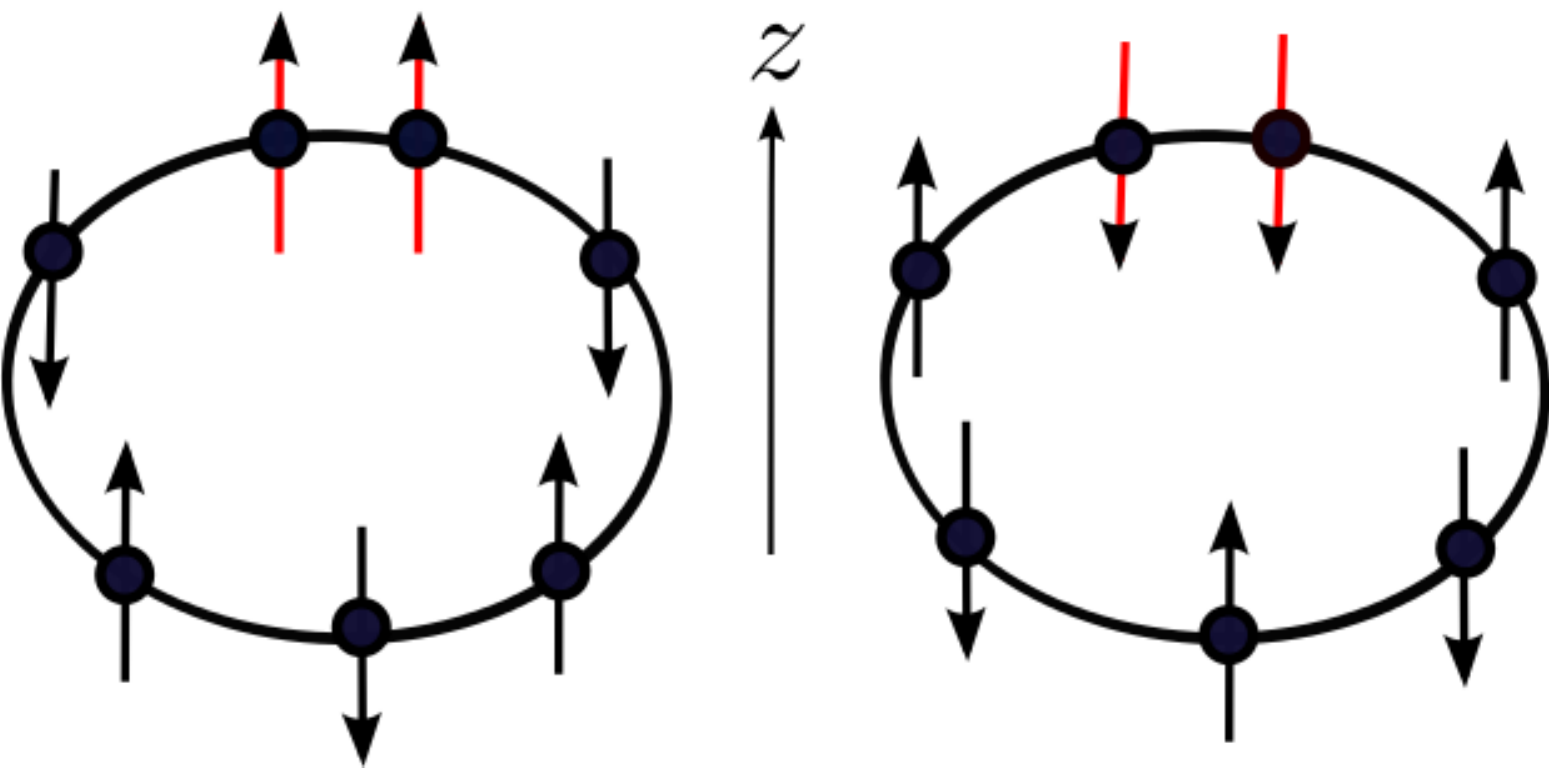}
\caption{Schematic sketch of the $2N$-fold classical degenerate ground states for antiferromagnetic molecular nanomagnets with odd-numbered coupled spins  in zero magnetic field.}
\label{ring1}
\end{figure}

In this Letter, we consider the general Hamiltonian of a cyclic molecular magnet, modeled as $N$ arbitrary large spins, regularly spaced on a circular ring \cite{ml2001}:
 \begin{align}
\hat H &=J\sum_{l=1}^{N} \hat{\bold{S}}_{l}\cdot\hat{\bold{S}}_{l+1} -D\sum_{l=1}^{N} \hat{S}_{l,z}^2  +g\mu_B  \bold{B} \cdot\sum_{l=1}^N \hat{\bold{ S}}_l.
\label{eqn4}
\end{align}
We assume periodic boundary condition $\hat{\bold{S}}_{N+1}=\hat{\bold{ S}}_1$ and the case of  odd $N$.  The first term represents the magnetic exchange coupling between spins with $J>0$, the second term is the  easy-axis magnetic anisotropy with $D>0$, and the last term is the Zeeman coupling. This model has a wide range of applications in many cases of physical interests.
For bipartite molecular wheels with even-numbered coupled spins, eq.\eqref{eqn4} is effectively a two-spin Hamiltonian \cite{mann}, which describes different molecular nanomagnets \cite{ml2001, og, da} depending on the values of $J$ and $D$. In the classical terminology, one describes such systems using spin coherent state (SCS) path integral formalism \cite{AB}, where the spins are assumed to be confined on a two-sphere $\mathcal{S}_2$. In this formalism, the spins can be thought of as classical vectors  aligned along the $z$-axis, but cant towards the direction of  the magnetic field.  The classical ground states consist of  two N\'eel vectors $\bold{n}_1$ and $\bold{n}_2$, which are degenerate (for $\bold{B}=0$) separated by a magnetic energy barrier of height $2DNs^2$; see Fig.\eqref{ring}.  When the barrier height is larger than the ground state energy,  the N\'eel vectors are well localized at the two degenerate minima of the magnetic potential; quantum tunneling lifts the degeneracy of the ground states and produces an energy splitting, which oscillates in the presence of a hard-axis magnetic field \cite{bar,ml2001} ({\it i.e.} $\bold{B}=B\bold{e}_x$).   This formalism has been successfully employed to describe quantum phenomena in physical molecules \cite{bar, ml2001} such as CsFe$_8$, with $N=8$, $s=5/2$; NaFe$_6$, with $N=6$, $s=5/2$, etc.  An exact numerical diagonalization \cite{cin} also corroborates the results of SCS path integral formalism. However, for molecular wheels with odd number of magnetic ions, SCS path integral is infeasible and the spin chain does not necessarily possess an effective two-spin Hamiltonian as the system is non-bipartite. In this case the classical ground state is frustrated and contains a topological soliton or domain wall, which is $ 2N$-fold degenerate as shown in Fig.\eqref{ring1}. The $N$-fold degenerate one-soliton states  with total $S_z=s$ can be written quantum-mechanically as
 \beq
 \ket{m_u}=\ket{\uparrow, \downarrow, \cdots,\uparrow,\downarrow,\underbrace{\textcolor{red}{\uparrow,\uparrow}}_{m,m+1},\downarrow,\uparrow\cdots ,\uparrow, \downarrow};
\label{up}
\eeq
where $m=1,\cdots,N$;   $\uparrow\equiv s$,  $\downarrow\equiv -s$, and the subscript $u$ denotes the up-up soliton.  The corresponding $N$-fold degenerate one-soliton states $\ket{m_d}$  with total $S_z=-s$ are obtained by $\pi$-rotation of all the spins in eq.\eqref{up}.

 In principle,  molecular wheels with even-numbered coupled spins should contain topological solitons as excited states, but they must come in pair by flipping an odd number of neighbouring spins from the fully anti-aligned N\'eel states. This has been demonstrated  in molecular clusters with a pure isotropic Heisenberg model via an exact numerical diagonalization \cite{enz1}. We will consider a hard-axis magnetic field along the plane of the magnet and the limit in which the exchange interaction is small compared to the anisotropy term, {\it i.e,} ($D\gg J$).  In this perturbative limit, we expect small quantum fluctuations stemming from the interaction and the magnetic field terms to lift these degenerate states;  this results in delocalization of the soliton with a formation of an energy band. The non-perturbative limit ($D\ll J$) is tricky and requires thorough investigation for finite odd-numbered rings as the semiclassical approach is infeasible.  For infinite chains, this limit corresponds to the Haldane limit \cite{dane}. Recently, there have been some studies on short odd-numbered magnetic rings with pure Heisenberg exchange interaction and open boundary condition \cite{cr}. To the best of our knowledge, no solution has been reported for short odd-numbered magnetic rings with periodic boundary condition in this non-perturbative limit. 

\textbf{Perturbation theory}--. Quantum dynamics in short antiferromagnetic molecular wheels (chains) are usually studied by semiclassical formalisms and exact numerical diagonalization techniques. However, the basic principles of perturbation theory apply to both classical and quantum systems; it does not rely on the length of the spins (large or small) or the size of the systems (finite or infinite). Thus, in many different occasions perturbation theory often reproduces the semiclassical results. We will follow the path of perturbation theory; it is customary to write eq.\eqref{eqn4} as a sum of three terms:
\bea
\hat H= \hat H_{0} +\hat {V}_1+\hat {V}_2.\eea
These terms are defined as
\begin{align}
\hat H_{0}&= -D\sum_{l=1}^{N} \hat{S}_{l,z}^2;\\ \hat{V}_1&= J\sum_{l=1}^{N}\lb \hat{S}_{l,z}\hat{S}_{l+1,z}+\frac{1}{2}(\hat{S}_{l}^+\hat{S}_{l+1}^-+\hat{S}_{l+1}^+\hat{S}_{l}^-)\rb;\\ \hat{V}_2&= \frac{h_x}{2}\sum_{l=1}^{N}\lb S_{l}^ + + \hat{S}_{l}^-\rb; \quad h_x=g\mu_B B;
\end{align}
where $\hat{S}_{l}^\pm= \hat{S}_{l,x}\pm i \hat{S}_{l,y} $ are the raising and lowering spin operators. The $N$-fold classical degenerate ground states are also degenerate ground states of $\hat H_0$:
\bea
\hat H_0\ket{m_u} = \mathcal{E}_0\ket{m_u}; 
\eea
where $\mathcal{E}_0= -DNs^2$. At first order in degenerate perturbation theory, we must diagonalize $\hat{V}$ in the degenerate subspace $\ket{m_u}$:
\begin{align}
\mathscr{H}^{1}_{m_u^\prime,m_u}=\braket{m_u^\prime|\hat V|m_u}\equiv\hat{V}_{m_u^\prime,m_u};\quad m_u^\prime,m_u=1,2,\cdots, N.
\label{firstor}
\end{align}
 The splitting terms (raising and lowering operators) in $\hat{V}=\hat {V}_1+\hat {V}_2$ map $\ket {m_u} $ to states out of the degenerate subspace;  consequently, the overlap between the resulting states and the degenerate states $\ket{m^{\prime}_u}$ vanishes. Thus, only the $S_z$ term in $\hat{V}_1$ gives a nonzero energy contribution and the eigenvalues are determined by the secular equation \cite{ll}:
\begin{align}
\text{det}(\hat{V}_{m^\prime,m}-\mathcal{E}^{1}\delta_{m^\prime,m})=0,
\label{firstdet}
\end{align}
where $\hat{V}_{m^\prime,m}=Js^2\lb 1-(N-1)\rb\delta_{m^\prime,m}$; hence,  the determinant in eq.\eqref{firstdet} is indeed diagonal, yielding $\mathcal{E}^m_{\pm}=\mathcal{E}^0+\mathcal{E}^{1} =-DNs^2-Js^2(N-2)$. Evidently, $\Delta=\mathcal{E}_{+}^m-\mathcal{E}_{-}^m=0$. 
A nonzero energy splitting is obtained when the transverse operators map the degenerate subspace into itself. This occurs at order $2s$; hence, $\Delta=\mathcal{E}_{+}^{(n<2s)}-\mathcal{E}_{-}^{(n<2s)}=0$. However, at order $2s$, we must consider small quantum fluctuation next to the position of the soliton. The small fluctuations(splitting) from $\hat{V}_1$ at order $2s$ give:
\begin{align}
&(S_{m+1}^-S_{m+2}^+)^{2s}\ket{m_u}\propto\ket{m_u+2};\nonumber \\&
(S_{m-1}^+S_{m}^-)^{2s}\ket{m_u}\propto\ket{m_u-2};
\label{so11}
\end{align}
where the states $\ket{m_u\pm 2}$ are defined periodically, thus they are elements of $\ket{m_u}$. However, fluctuations from $\hat{V}_2$ at order $2s$ map to a state with total $S_z=-s$:
\begin{align}
&(S_{m}^-)^{2s}\ket{m_u}\propto\ket{m_d-1}; \thinspace
(S_{m+1}^-)^{2s}\ket{m_u}\propto\ket{m_d+1}; 
\label{so12}
\end{align}
where
 \beq
\ket{m_d-1}\equiv\ket{\tilde{m}_d}=\ket{\uparrow, \downarrow, \cdots,\downarrow,\uparrow,\underbrace{\textcolor{red}{\downarrow,\downarrow}}_{m-1,m},\uparrow,\downarrow\cdots ,\uparrow, \downarrow}
\label{down}.
\eeq
 It is evident that fluctuations away from the soliton, such as $(S_{m+4}^+S_{m+5}^-)^{2s}$ and $ (S_{m+2}^+S_{m+3}^-)^{2s}$, produce states with higher energy, which give small corrections in the perturbative series, thus can be ignored. 

Degenerate perturbation theory has been developed for single molecule magnets \cite{gga,chud10}. We employ this formalism to  our problem; the matrix elements for the present problem are given by
\begin{align}
\mathscr{H}_{m^{\prime}_u,m_u}= \braket{m^{\prime}_u|\hat{V}_1\mathcal{R}_1^{{2s-1}}|m_u},
\label{mat11}
\end{align}
\begin{align}
 \mathscr{H}_{\tilde{m}_d^{\prime},m_u}=\braket{\tilde{m}^{\prime}_d|\hat{V}_2\mathcal{R}_2^{{2s-1}}|m_u},
\label{matr}
\end{align}
where 
 \bea \mathcal{R}_j^{{2s-1}}= \lb\frac{\mathcal{P}_c}{\mathcal{E}_{0}-\hat H_{0}}\hat{V}_j\rb^{2s-1};\quad j=1,2; \eea 
  is the $(2s-1)$-th order transition matrix   and $\mathcal{P}_c=1-\sum_{m=1}^{N} \ket{m}\bra{m}$ is the complementary projection operator. The total matrix elements can be written as
\bea
\mathscr{H}^{2s}=\mathscr{H}_{m^{\prime}_u,m_u}+\mathscr{H}_{\tilde{m}^{\prime}_d,m_u}.
\eea
\textbf{Energy band}--. 
The diagonalization of the $N\times N$ matrix in the preceding section gives the energy band (spliting) of the system.  The components of this matrix can indeed be computed exactly for the one-soliton states. The only terms that survive in eqs.\eqref{mat11} and \eqref{matr} are indeed the fluctuations next to the soliton, as defined in  eqs.\eqref{so11} and \eqref{so12}.  Thus,
\begin{align}
\mathscr{H}_{m^{\prime}_u,m_u}&=  \braket{m^{\prime}_u|\lb \mathcal{T}_{m+1,m+2}+\mathcal{T}_{m-1,m}\rb |m_u};
\label{matr1}\\
\mathscr{H}_{\tilde{m}^{\prime}_d,m_u}&= \braket{\tilde{m}_d^{\prime}|\lb \mathcal{T}_m +\mathcal{T}_{m+1}\rb|m_u},
\label{matr11}
\end{align}
where the $\mathcal{T}$ operators are defined  in the Appendix.
The matrix elements are found to be:
\begin{align}
\mathscr{H}^{2s}&= \mathscr{C}_{J}[\delta_{m^{\prime}_u,m_u+2}+\delta_{m^{\prime}_u,m_u-2}] \nonumber\\&+ \mathscr{C}_{h_x}[\delta_{\tilde{m}^{\prime}_d,\tilde{m}_d+1}+\delta_{\tilde{m}_d^{\prime},\tilde{m}_d-1}],
\label{hop3}
\end{align}

where the $\mathscr{C}'s$ are given by
\begin{align}
\mathscr{C}_{J}&=\pm\lb\frac{J}{2}\rb^{2s}\prod_{\sigma=1}^{2s} \sigma(2s-\sigma+1)\prod_{\sigma=1}^{2s-1}\frac{1}{2D\sigma(2s-\sigma)},\label{prod}\\&=\pm 8Ds^2\lb\frac{J}{4D}\rb^{2s}.\\
\mathscr{C}_{h_x}&=\pm\lb\frac{h_x}{2}\rb^{2s}\prod_{\sigma=1}^{2s} \sqrt{\sigma(2s-\sigma+1)}\prod_{\sigma=1}^{2s-1}\frac{1}{D\sigma(2s-\sigma)},\label{prod1}\\&=\pm \frac{2Ds}{(2s-1)!}\lb\frac{h_x}{2D}\rb^{2s}\sim \pm D\sqrt{\frac{s}{\pi}}\lb\frac{eh_x}{4Ds}\rb^{2s}.
\label{cc}
\end{align}

The last result in eq.\eqref{cc} follows from the Stirling approximation $n!\sim \sqrt{2\pi n}\lb\frac{n}{e}\rb^n$ for $s\gg 1$. The first two products in eqs.\eqref{prod}  and \eqref{prod1} emanate from the relation of the raising and lowering operators. The second products stem from the energy denominators. There is a  plus or minus sign in these results which can be understood as follows. There are $2s-1$ negative energy denominators in eqs.\eqref{matr1} and \eqref{matr11}. Thus, if $s$ is integer, $2s-1$ is odd, and  a minus sign appears; whereas for half-odd integer spins, $2s-1$ is even, and  a plus sign appears.  Due to the cyclic symmetry of the ring, 
in  matrix representation  we obtain  a circulant matrix \cite{dav}: 
\begin{equation}
\mathscr{H}^{2s} =
 \begin{bmatrix}
  0 & \mathscr{C}_{h_x} & \mathscr{C}_{J}&0& \cdots & \mathscr{C}_{J}& \mathscr{C}_{h_x} \\
  \mathscr{C}_{h_x} & 0 &\mathscr{C}_{h_x}&\mathscr{C}_{J}&\cdots & 0&\mathscr{C}_{J} \\
  \mathscr{C}_{J} &  \mathscr{C}_{h_x} &  0 & \mathscr{C}_{h_x}& \mathscr{C}_{J}&\cdots&0 \\
  \vdots  & \mathscr{C}_{J} &\mathscr{C}_{h_x}& \ddots&\ddots&\ddots& \vdots  \\
  0  & 0 &\ddots& \ddots&\ddots&\ddots& \mathscr{C}_{J}  \\
  \mathscr{C}_{J} & \cdots &\ddots & \ddots &\mathscr{C}_{h_x} &0&\mathscr{C}_{h_x} \\
 \mathscr{C}_{h_x} & \mathscr{C}_{J} & 0&\cdots  &\mathscr{C}_{J} &\mathscr{C}_{h_x}&0
 \end{bmatrix}.
 \label{eqn28}
\end{equation}
  
 The eigenvalues are given by the finite cyclic group  Fourier transform \cite{dav}:
 \bea
 \epsilon_m= \sum_{j=0}^{N-1}c_j\tau_{m}^j,
 \eea
where  $\tau_{m}^j= e^{i\lb 2\pi/N\rb j m}$,  $\thinspace m=0,1, 2,\cdots,N-1$ modulo $N$ is the  shift quantum number  \cite{enz},  and $c_m$ are the row or the column entries of $\mathscr{H}^{2s}$. The  corresponding normalized eigenvector are column vectors with components  \bea v_m^j=\frac{\tau_{m}^j}{\sqrt{N}}.\eea

 The one-soliton energy band for our model then simplifies to
\begin{align}
\epsilon_m& = \mathscr{C}_{J}(\tau_{m}^2 +\tau_{m}^{N-2})+\mathscr{C}_{h_x}(\tau_{m} +\tau_{m}^{N-1}),\nonumber\\&
=2\mathscr{C}_{J}\cos\lb\frac{4\pi m}{N}\rb+2\mathscr{C}_{h_x}\cos\lb\frac{2\pi m}{N}\rb.
\label{eqn31}
\end{align}
It should be noted that the result of degenerate perturbation theory is unchanged if one uses the normalized linear superposition of the $2N$-fold degenerate one-soliton states $\ket{m_s}=\frac{1}{\sqrt{2}}\lb\ket{m_u}+\ket{m_d}\rb$  with $S_z=0$.

\textbf{Chiral States and Polarized Neutron Scattering}.
 In terms of the momentum $q= 2\pi m /aN$, the energy bands can be written as
\begin{align}
 \epsilon_q  
=-2\cos(2\pi s)[|\mathscr{C}_{J}|\cos\lb{2qa}\rb + |\mathscr{C}_{h_x}|\cos\lb{qa}\rb].
\label{soloen}
\end{align}

For spin-$\frac{1}{2}$ systems such as CsCoBr$_3$ and CsCoCl$_3$  with odd number of sites, the energy band reduces to a well-known Villain result \cite{nat, vill}. This is expected as large spin systems should in principle reproduce the results for small spin systems at least in perturbation theory. The major difference, however, is  the size of the systems, $N$. Conventional quantum spin systems usually have large number of sites (thermodynamic limit) and strong quantum fluctuations as opposed to molecular nanomagnets. As a consequence, spin systems with finite $N$ and large spins usually behave as classical systems but possess quantum phenomena \cite{mann}.  The one-soliton energy band in eq.\eqref{soloen} is thus the general formula for any system size  and arbitrary spins. The argument $qa$ and $2qa$ signify  the total hopping sites of the soliton, as a result of small quantum fluctuations.
For a single large spin, $J=0$, eq.\eqref{eqn4} reduces to $\hat H= -DS_z^2+h_xS_x$, which describes Mn$_{12}$Ac; then eq.\eqref{soloen} becomes the exact ground state tunneling splitting \cite{gga} with $q=0$, which corresponds to the transition $\ket{\uparrow}\leftrightarrow\ket{\downarrow}$. In order to characterize states, it is expedient to introduce the vector chiral operator defined as \cite{nat} $\bold K= \sum_l \bold{S}_l\times \bold{S}_{l+1}$. The transverse component can be diagonalized in the one-soliton states $\ket{m_s}$, where $\ket{m_s}$ is the linear superposition of the  $2N$-fold degenerate one-soliton states.  We find that $\mathcal{K}^x=\braket{\tilde{m}_s^\prime|(K_x)^{2s}|m_s}$ is given by
\bea
\mathcal{K}^x= \mathscr{K}^s\sin(qa),
\eea
where  $\mathscr{K}^s=2(2s)!(s)^{2s}$. As shown in  Fig.\eqref{band}, for half-odd integer spins, the soliton ground state is chiral located at $q=\pm\frac{\pi}{2a}$, while for integer spins  the state at $q=0$ is the non-chiral soliton ground state.

Due to chirality the quantity of interest is the polarized neutron scattering intensity, which is an indispensable process in experiments \cite{moon}. For spin $1/2$ systems the emergence of soliton in quantum antiferromagnetic spin chain has been studied experimentally using polarized neutron scattering techniques \cite{nat}. In the present problem, experimental measurement should be reminiscent of that of spin $1/2$ systems \cite{nat}. The crucial difference, however, is that in the present problem the energy band is formed at order $2s$ and the system size is  finite and short. The quantity of interest is usually the magnetic scattering intensity, which  has the form \cite{nat, moon}:
\begin{figure}[ht]
\centering
\includegraphics[scale=0.3]{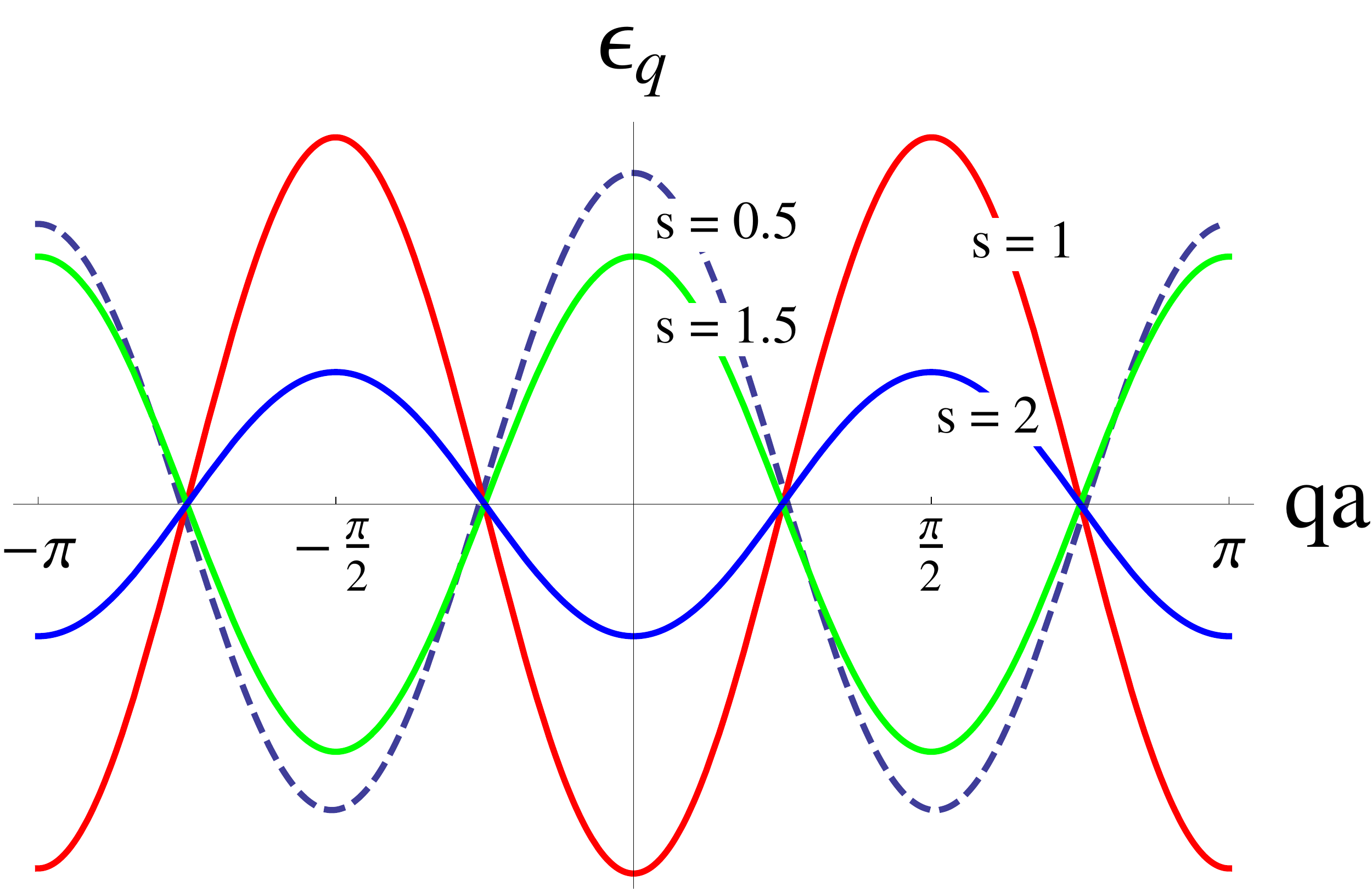}
\caption{The one-soliton energy band for different values of the spins with $J/4D=0.3$ and $h_x/2D=0.05$. The ground state for half-odd integer spins is chiral located at $q=\pm \pi/2a $; whereas that of integer spins is non-chiral located at $q=0$. As the spin increases the energy band approaches a constant value.}
\label{band}
\end{figure}

\begin{align}
\lb\frac{d^2\sigma}{d\omega d\Omega}\rb_{\pm}\propto \sum_{\alpha\beta\gamma}[\delta_{\alpha\beta}\mathcal{S}^{\alpha\beta}(\bold{Q},\omega) \pm i\mathcal{K}^{\gamma}(\bold{Q},\omega)],
\label{chiral}
\end{align}
where  $\bold{Q}= \bold{q}-\bold{q}^\prime$ is the momentum transfer, $\omega=E-E^\prime$ is the energy transfer, and $\alpha,\beta,\gamma =x,y,z$. The plus or minus sign denotes the directions of the magnetic field. The magnetic dynamical structure factor $\mathcal{S}^{\alpha\beta}(\bold{Q},\omega)$  and the dynamical chirality $\mathcal{K}^{\gamma}(\bold{Q},\omega)$ are given by
\begin{align}
\mathcal{S}^{\alpha\beta}&=\int_{-\infty}^{\infty} dt\braket{S_{-\bold{Q}}^{\perp\alpha}(0)S_{\bold{Q}}^{\perp\beta}(t)}e^{-i\omega t}\label{inela}
\\ \mathcal{K}^{\gamma}&=\epsilon_{\alpha\beta\gamma}\mathcal{S}^{\alpha\beta}.
\label{inela1}
\end{align}
where $S_{\bold{Q}}^{\perp\alpha}= S_{\bold{Q}}^{\alpha}-\hat{Q}_\alpha\sum_\sigma S_{\bold{Q}}^{\sigma}\hat{Q}_\sigma$ is the molecular spin component transverse to the momentum transfer and $\epsilon_{\alpha\beta\gamma}$ is the antisymmetric Levi-Civita tensor. In the absence of chirality,  eq.\eqref{chiral} reduces to usual formula for unpolarized neutrons scattering \cite{og,love}. The thermal average in eq.\eqref{inela} must be performed in the one-soliton states $\ket q$, which represents coherent superpositions of solitons at all the magnetic spin sites. As shown in eqs.\eqref{so11} and \eqref{so12}, a non-zero correlation function is obtained by taking the transverse components in eqs.\eqref{inela} and \eqref{inela1}  to order $2s$. This is the case for every quantity that requires the computation of correlation functions such as  the spin-lattice relaxation time. Without computing the neutron scattering intensity explicitly, we can get an intuitive understanding of its structure.  Depending on the strengths of the magnetic field and the coupling constants $J$ and $D$, the energy band $\epsilon_q$ may increase or decrease as $s$ increases (see Fig.\eqref{band}). Thus, it is evident that the scattering intensity will have a similar trend but approaches a constant value at very large $s$.

\textbf{Conclusion--}. In conclusion, we have studied an antiferromagnetic molecular nanomagnet with an odd-numbered coupled spins,  in the presence of a transverse magnetic field applied along the hard-axis anisotropy. We considered arbitrary spin with periodic boundary condition. In this case we showed  that the N\'eel state is frustrated, resulting in a defect which introduces a topological soliton. As the soliton can be placed anywhere along the cyclic chain, the resulting energy is $N$-fold degenerate in the sector with a total $S_z=s$. Small quantum fluctuation stemming from the interaction term translates the soliton by two magnetic sites in the same state, whereas fluctuation from the magnetic field term translates the soliton by one magnetic site in a different state with total $S_z=-s$. Delocalization of the soliton by quantum fluctuation forms an energy band  at order $2s$ in degenerate perturbation theory. We obtained the energy band and showed that the soliton ground state is chiral for half-odd integer spins and non-chiral for integer spins. We  further demonstrated the structure of the inelastic polarized  neutron scattering intensity, which should be of experimental interest. Our results are general and apply to systems with arbitrary  size and spins.

\acknowledgments
The authors would like to thank  NSERC of Canada for financial support. We are also grateful to Juergen Schnack for useful discussions.  This work started at Universit\'e de Montr\'eal and was completed at Perimeter Institute for Theoretical Physics.

\textbf{Appendix}--.
\label{apen}
The $\mathcal{T}$ operators in eqs.\eqref{matr1} and \eqref{matr11} are given by
\begin{align}
\mathcal{T}_{m+1,m+2}=&\lb\frac{J}{2}\rb^{2s}S_{m+1}^-S_{m+2}^+\lb\frac{\mathcal{P}_c}{\mathcal{E}_{0}-\hat H_{0}}S_{m+1}^-S_{m+2}^+\rb^{2s-1},\label{ap1}\\
\mathcal{T}_{m-1,m}=&\lb\frac{J}{2}\rb^{2s}S_{m-1}^+S_{m}^-\lb\frac{\mathcal{P}_c}{\mathcal{E}_{0}-\hat H_{0}}S_{m-1}^+S_{m}^-\rb^{2s-1},\label{ap2}\\
\mathcal{T}_{m}=& \lb\frac{h_x}{2}\rb^{2s}S_{m}^-\lb\frac{\mathcal{P}_c}{\mathcal{E}_{0}-\hat H_{0}}S_{m}^-\rb^{2s-1}.
\label{ap3}
\end{align}

\end{document}